\documentclass[9pt,twocolumn,twoside]{opticajnl}
\journal{opticajournal} 

\setboolean{shortarticle}{true}

\usepackage{lineno}

\usepackage{graphicx}
\usepackage{dcolumn}
\usepackage{bm}
\usepackage{color}
\usepackage{amsfonts}

\title{$\mathcal{PT}$-symmetric photonic lattices with type-II Dirac cones}

\author[1]{Qian Tang}
\author[2]{Milivoj R. Beli\'c}
\author[3]{Hua Zhong}
\author[3]{Meng Cao}
\author[3]{Yongdong Li}
\author[3,*]{Yiqi Zhang}

\affil[1]{Ministry of Education Key Laboratory for Nonequilibrium Synthesis and Modulation of Condensed Matter, Shaanxi Province Key Laboratory of Quantum Information and Quantum Optoelectronic Devices, School of Physics, Xi'an Jiaotong University, Xi'an 710049, China}
\affil[2]{Division of Arts and Sciences, Texas A\&M University at Qatar, P.O. Box 23874 Doha, Qatar}
\affil[3]{Key Laboratory for Physical Electronics and Devices, Ministry of Education, School of Electronic Science and Engineering, Xi'an Jiaotong University, Xi'an 710049, China}

\affil[*]{zhangyiqi@xjtu.edu.cn}

\begin{abstract}
	The type-II Dirac cone is a special feature of the band structure, whose Fermi level is represented by a pair of crossing lines.
	It has been demonstrated that such a structure is useful for investigating topological edge solitons, and more specifically, for mimicking the Kline tunneling.
	However, it is still not clear what the interplay between type-II Dirac cones and the non-Hermiticity mechanism will result in.
	Here, this question is addressed; in particular, we report the $\mathcal{PT}$-symmetric photonic lattices with type-II Dirac cones for the first time.
	We identify a slope-exceptional ring and name it the \textit{type-II exceptional ring}.
	We display the restoration of the $\mathcal{PT}$ symmetry of the lattice by reducing the separation between the sites in the unit cell.
	Curiously, the amplitude of the beam during propagation in the non-Hermitian lattice with $\mathcal{PT}$ symmetry only decays because of diffraction,
	whereas in the $\mathcal{PT}$ symmetry-broken lattice it will be amplified, even though the beam still diffracts.
	This work establishes the link between the non-Hermiticity mechanism and the violation of Lorentz invariance in these physical systems.
\end{abstract}

\setboolean{displaycopyright}{false} 

\begin{document}
	
	\maketitle

	The $\mathcal{PT}$ symmetry was first proposed in Quantum Mechanics in 1998~\cite{bender.prl.80.5243.1998}, on the account of demonstrating a curious new feature,
	that some $\mathcal{PT}$-symmetric non-Hermitian Hamiltonians may still display entirely real spectra of eigenvalues. This striking discovery led fast to the formulation of a new version of Quantum Mechanics, the Non-Hermitian Quantum Mechanics.
	In the past few decades, the concept attracted a lot of attention from researchers in various fields, but especially from the optical and photonic communities,
	since it is quite natural and relatively simple to realize $\mathcal{PT}$-symmetric structures in optical systems~\cite{ganainy.ol.32.2632.2007,makris.prl.100.103904.2008,musslimani.prl.100.030402.2008,guo.prl.103.093902.2009,ruter.np.6.192.2010,regensburger.nature.488.167.2012,chang.np.8.524.2014,zhang.prl.117.123601.2016}, including waveguide arrays, optical microresonators, photonic crystals, and so on. 
	It is worth mentioning that the $\mathcal{PT}$ symmetry in optics~\cite{cham.np.11.799.2015} belongs to the well-known quantum-optical analogy~\cite{longhi.lpr.3.243.2009}.
	In this limited space we cannot include much of the vast literature on $\mathcal{PT}$ symmetry; we only recommend some recent reviews~\cite{konotop.rmp.88.035002.2016,longhi.epl.120.64001.2018,ganainy.np.14.11.2018,miri.science.363.eaar7709.2019,ashida.aip.69.249.2020,gupta.am.32.1903639.2020,wang.aop.15.442.2023}.
	
	In photonic lattice systems, especially lattices with Dirac cones, the  $\mathcal{PT}$ symmetry is very important~\cite{szameit.pra.84.021806.2011,zhang.rrp.68.230.2016,wang.josab.40.1443.2023},
	owing to the tremendous applicative potential of combined non-Hermitian properties and topological photonics~\cite{ozawa.rmp.91.015006.2019,parto.nano.10.403.2021,wang.jo.23.123001.2021,li.nn.18.706.2023,nasari.ome.13.870.2023,yan.nano.12.2247.2023}.
	One fascinating feature of non-Hermitian photonic lattices is that the bands may coalesce into exceptional rings, which are different from the 
	diabolic points (e.g., Dirac points) in the Hermitian systems~\cite{li.nn.18.706.2023}.
	Both the eigenstates and eigenvalues on the exceptional ring (which is a ring of exceptional points) are degenerate; but, only the eigenvalues at the diabolic points are degenerate.
	Another convenient feature of the non-Hermitian photonic lattices is that they may possess totally real eigenvalues~\cite{szameit.pra.84.021806.2011,weimann.nm.16.433.2016,xia.science.372.72.2021},  provided proper operations are performed on the lattice.
	
	On the other hand, the type-II diabolic point is quite rare in comparison with its type-I counterpart~\cite{rechtsman.nature.496.196.2013,jena.jpd.57.305101.2024},
	because the Fermi surface corresponding to the former case is a pair of crossing lines, while for the latter it is a point~\cite{milicevic.prx.9.031010.2019}.
	In addition, the quasi-particles corresponding to the type-II diabolic points violate the Lorentz invariance~\cite{xu.sa.3.e1603266.2017,yang.nc.8.257.2017,noh.prl.119.016401.2017,fei.prb.96.041201.2017}, which is a feature worthwhile investigating. Also interesting  would be to construct other lattices possessing type-II Dirac points. Namely,  
	there is only one simple lattice that possesses type-II Dirac points naturally~\cite{jin.aqt.3.2000015.2020,zhong.ap.3.056001.2021,tian.aop.17.53503.2022,tian.oe.31.20812.2023}.
	It is not an easy task to seek for other lattices with band structures supporting type-II Dirac points,
	because one has to exert carefully chosen deformations on the lattice sites, to effect such a band~\cite{pyrialakos.prl.119.113901.2017,pyrialakos.nc.11.2074.2020,hu.prl.121.024301.2018,wu.prl.124.075501.2020,milicevic.prx.9.031010.2019,mann.nc.9.2194.2018,noh.np.13.611.2017,yang.nc.8.97.2017}.
	This makes the investigation of the interplay between type-II Dirac points and non-Hermitian mechanisms an extraordinarily difficult task. 
	
	In this Letter, based on the lattice with a natural type-II Dirac point, which is unique for investigating the aforementioned interplay, 
	we report the \textit{type-II exceptional ring} and show the establishment of $\mathcal{PT}$ symmetry for the first time.
	The reason for introducing the \textit{type-II exceptional ring} is that it originates in the interplay between type-II diabolic points and the non-Hermiticity mechanism.
	The exceptional point associated with type-I diabolic point is the type-I exceptional point,
	which is common and nothing new.
	We will also show that the non-Hermitian lattice can be adjusted to be $\mathcal{PT}$-symmetric,
	by suppressing the separation among the three sites in each unit cell.
	At last, the $\mathcal{PT}$ symmetry of the lattice is demonstrated by comparing the discrete diffraction of Gaussian beams launched into properly prepared conservative and non-Hermitian lattices.
	The Gaussian beam is always damped if the $\mathcal{PT}$ symmetry is satisfied, but it will be amplified if the $\mathcal{PT}$ symmetry is broken.

	The propagation dynamics of a light beam can be described by the normalized nonlinear Schr\"odinger-like paraxial wave equation:
	\begin{align} \label{eq1}
		i \frac{\partial \psi}{\partial z} = -\frac{1}{2}\nabla^2 \psi -[ \mathcal{R}_{\rm re}(x,y) + i \mathcal{R}_{\rm im}(x,y)] \psi - |\psi|^2 \psi ,
	\end{align}
	where ${\nabla^2=\partial_x^2+\partial_y^2}$ is the transverse Laplacian,
	$x$ and $y$ are the spatial
	coordinates normalized to $r_0$, which is the characteristic transverse scale, and 
	$z$ is the propagation distance normalized to the diffraction length $kr_0^2$, in which
	${k = 2 n_0 \pi/\lambda }$ is the wave number, 
	$n_0$ is the ambient refractive index,
	and $\lambda$ is the wavelength.
	The function $\mathcal{R}_{\rm re} $ stands for the profile of the potential landscape (the refractive index modulation) that can be expressed using the Gaussian functions:
	$
	{\mathcal{R}_{\rm re}  = p_{\rm re} \sum_{m,n} \exp \left\{ - [(x - x_{m,n})^2 + (y - y_{m,n})^2]/{\sigma^2} \right\} ,}
	$
	with $(x_{m,n},y_{m,n})$ being the grid coordinates of the lattice.
	
\begin{figure}[htbp]
		\centering
		\includegraphics[width=0.9\columnwidth]{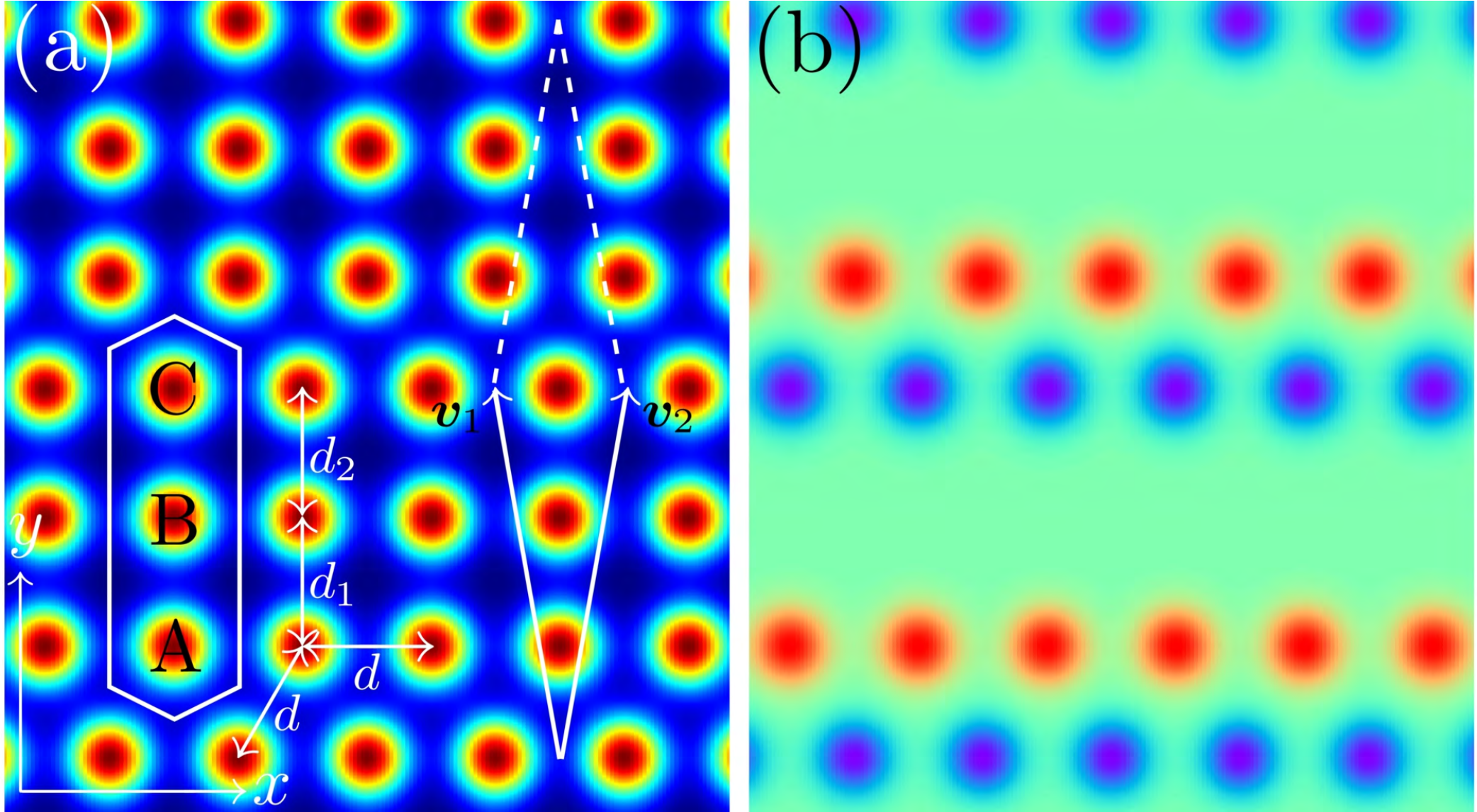}
		\caption{(a) Photonic lattice with type-II Dirac cones. There are three sites (A, B, C) in one unit cell, as indicated by a white hexagon. The lattice constant is $d$, and the separations of the three sites in each unit cell are adjusted by $d_1$ and $d_2$.  The basis vectors of the Bravais lattice are 
			${\bm{v}_1 = [- d/2,2d + \sqrt{3}d/2]}$ and ${\bm{v}_2 = [d/2,2d + \sqrt{3}d/2]}$. 
			(b) Imaginary part of the lattice, with site B being conservative, site A gainy, and site C lossy.}
		\label{fig1}
	\end{figure}

	In Fig.~\ref{fig1}(a), we display the lattice with type-II Dirac cones in its band structure,
	in which $d$ is the fixed lattice constant, equal to the separation between two neighboring unit cells along the $x$ direction, and
	${p_{\rm re} = k^2 r_0^2 \delta n_{\rm re} / n_0}$ is the depth of the waveguide, 
	with $\delta n_{\rm re}$ being the refractive index contrast. Parameter $\sigma$ determines the width of the waveguide.
	In the unit cell we fix site B, while sites A and C  can be shifted horizontally. 
	The separation between sites B and A is denoted by $d_1$ and that between sites B and C is denoted by by $d_2$. 
	In this work we assume that ${d_1=d_2}$, which makes the analysis simpler. 
	The function $\mathcal{R}_{\rm im} $ in Eq.~(\ref{eq1}) represents the imaginary part of the potential, which can also be expressed using Gaussian functions, with
	$p_{\rm im}$ being the strength of the potential.
	Figure~\ref{fig1}(b) gives the profile of $\mathcal{R}_{\rm im} $ that will be considered in this work---there is loss on site A and gain on site C, but there is neither gain nor loss on site B.
	
	For the convenience of analysis, we write the Hamiltonian of the system by using the tight-binding method, with the nearest-neighbor coupling being considered only,
	as follows:
	
	\begin{equation}\label{eq2}
		\mathcal{H}=
		\begin{bmatrix}
			H_{11}+ i\gamma & w & H_{13} \\
			w & H_{11} & w \\
			H_{13}^* & w & H_{11} - i\gamma
		\end{bmatrix},
	\end{equation}
	where $w$ is the intra-cell coupling strength, $n$ is the inter-cell coupling strength, and
	$H_{11}=2n\cos [{\bm k}\cdot ({\bm v}_2-{\bm v}_1)]$, $H_{13}=n [\exp(-i{\bm k}\cdot {\bm v}_1)+\exp(-i{\bm k}\cdot {\bm v}_2)]$ are the diagonal and off-diagonal components, respectively. Furthermore,
	${{\bm k} = [k_x, k_y]}$ is the Bloch momentum,
	and $\gamma$ weights the on-site loss or gain.
	
\begin{figure*}[t]
		\centering
		\includegraphics[width=0.98\textwidth]{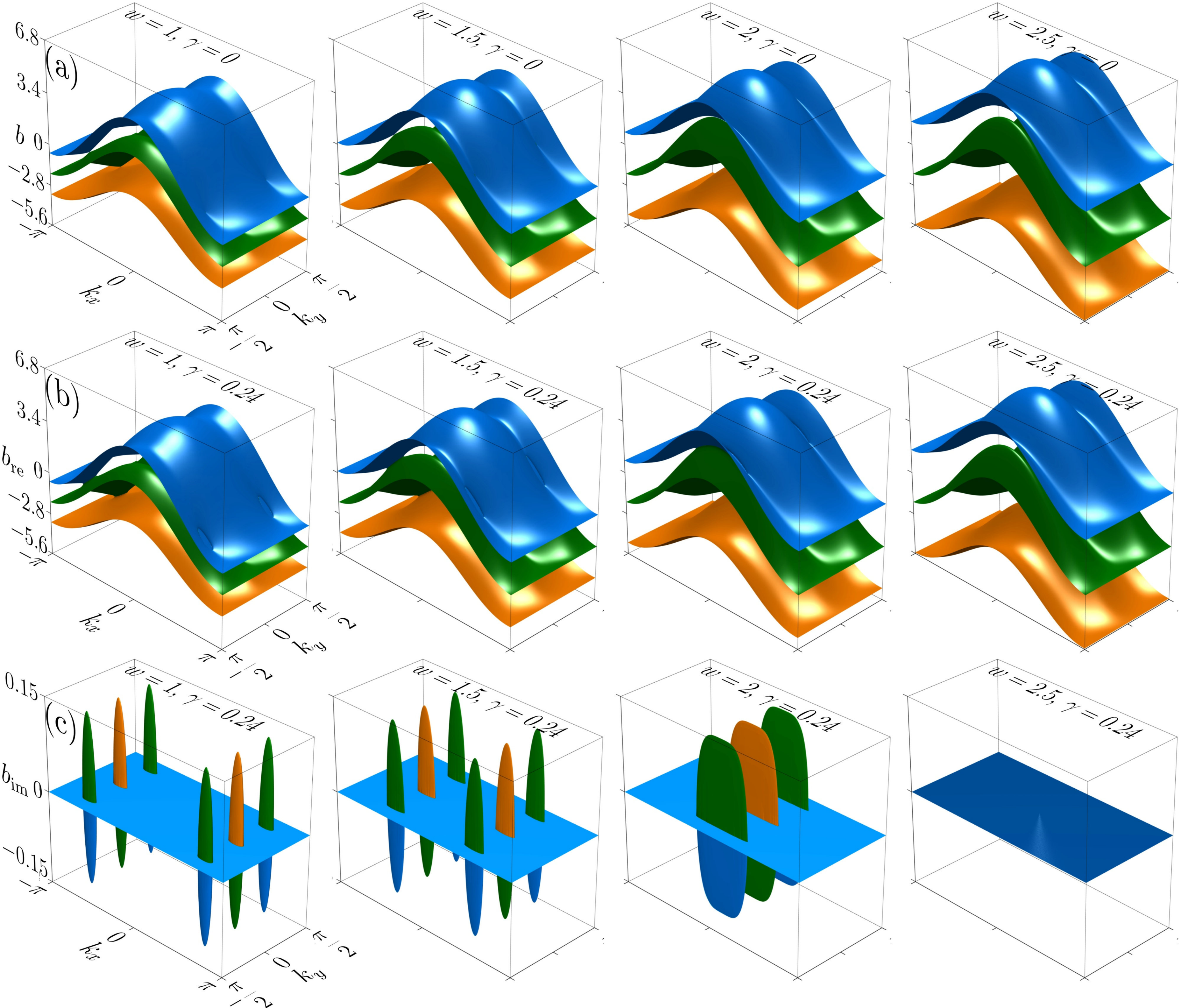}
		\caption{Band structure of the non-Hermitian lattice with type-II Dirac cones, based on the tight-binding method under the condition ${w\ge n}$.
			(a) Conservative case, with $w$ increasing gradually.
			(b) Real part of the band structure, with $w$ increasing gradually.
			(c) Imaginary part of the band structure corresponding to the real part in (b).
			Other parameters:  ${n=1}$ and ${\gamma=0.24}$. }
		\label{fig2}
	\end{figure*}

	Firstly, we consider the case with the intra-cell coupling being not smaller than the inter-cell coupling strength: ${w\ge n}$,
		which corresponds to ${d_1=d_2\le d}$ in Fig.~\ref{fig1}(a).
	For this case, we set ${n=1}$, so that only $w$ need to be adjusted.
	In Fig.~\ref{fig2}(a), we show the band structures of the conservative lattices with ${\gamma=0}$,
	to be compared with those of the non-Hermitian lattices.
	The band structures are  obtained numerically, by diagonalizing the Hamiltonian in Eq.~(\ref{eq2}), which is a function of $k_x$ and $k_y$.
	The lattice is regular if ${w=1}$, and the band structure possesses type-II Dirac cones between neighboring bands~\cite{jin.aqt.3.2000015.2020,zhong.ap.3.056001.2021,tian.aop.17.53503.2022,tian.oe.31.20812.2023}.
	As $w$ increases, the Dirac cones move towards ${k_x=0}$, which results in their meeting and coalescing with each other (see the band structure for ${w=2}$), 
	and disappearing ultimately, to form band gaps (see the band structure for ${w=2.5}$).
	
	Corresponding to the conservative case in Fig.~\ref{fig2}(a), 
	we show the band structures of the non-Herimitian lattices with ${\gamma=0.24}$ in Figs.~\ref{fig2}(b) and \ref{fig2}(c),
	which include the real and imaginary parts, respectively.
	Different from the conservative lattice, the type-II Dirac cones in the band structure are replaced by the exceptional elliptic rings.
	In the $\mathcal{PT}$-symmetric honeycomb lattice~\cite{szameit.pra.84.021806.2011}, 
	the exceptional rings are circular and parallel to the Bloch momentum plane.
	However, the exceptional rings here are neither parallel nor orthogonal to the Bloch momentum plane,
	as shown in panels with ${w=1}$ and ${w=1.5}$ in Fig.~\ref{fig2}(b).
	Since the exceptional ring is inherited from the type-II Dirac cone, it is reasonable to call it the \textit{type-II exceptional ring}.
	For the case with ${w=2}$, the type-II exceptional rings that are symmetric about ${k_x=0}$ coalesce, to form a curved ring,
	which is also symmetric about ${k_x=0}$.
	Similar to the conservative case in Fig.~\ref{fig2}(a), the exceptional rings disappear if ${w}$ increases further,
	as shown in the panel with ${w=2.5}$ in Fig.~\ref{fig2}(b).
	It is worth mentioning that the band structures with ${w=2.5}$ in Figs.~\ref{fig2}(a) and \ref{fig2}(b) are the same,
	and the reason is that the $\mathcal{PT}$ symmetry condition is satisfied and the eigenvalues of the Hamiltonian in Eq.~(\ref{eq2}) are completely real,
	as depicted by the imaginary part of the band structure with ${w=2.5}$ in Fig.~\ref{fig2}(c).
	For other cases with exceptional rings in the real part of the band structure,
	there are imaginary eigenvalues in the regions where the exceptional rings exist.
	In Fig.~\ref{fig2}(c), the imaginary part of the band structure with increasing intra-cell coupling strength $w$ is exhibited, to 
	display the realization of the $\mathcal{PT}$ symmetry.
	Following the same procedure, we fix the intra-cell coupling strength ${w=1}$ and increase the inter-cell coupling strength $n$ gradually.
	The corresponding band structures can be found in the \textbf{Supplemental Material}. They do not display $\mathcal{PT}$ symmetry.
		That is, the $\mathcal{PT}$ symmetry condition can be fulfilled via a deformation operation, by making the intra-cell coupling stronger.
	
\begin{figure}[t]
		\centering
		\includegraphics[width=0.9\columnwidth]{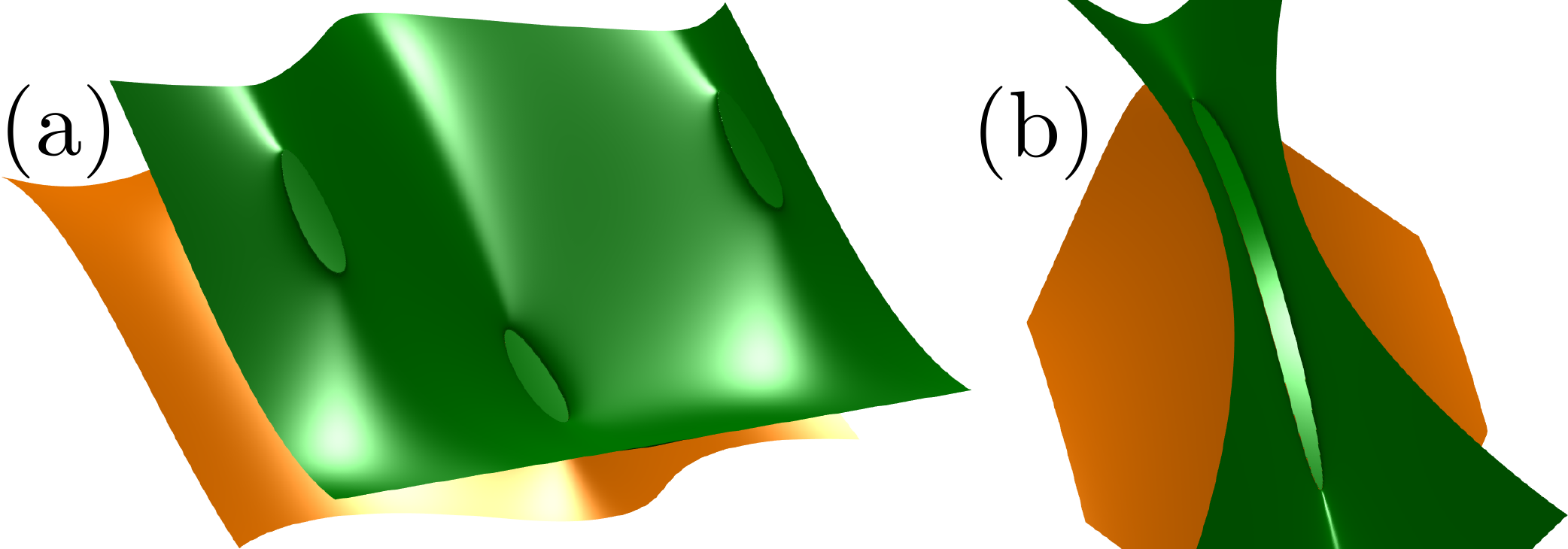}
		\caption{Magnified exceptional rings. (a) ${w=n=1}$. (b)  ${n=1}$ and ${w=2}$. Other parameters are same as those in Fig.~\ref{fig2}(b).}
		\label{fig4}
	\end{figure}

	In the previously reported exceptional rings, the eigvalues around the ring are all the same~\cite{szameit.pra.84.021806.2011,cerjan.np.13.623.2019,liu.prl.129.084301.2022,li.nn.18.706.2023},
	however this is not true for the type-II exceptional rings.
	In Fig.~\ref{fig4}, we show the magnified type-II exceptional rings of Fig.~\ref{fig2}(b), with ${w=1}$ and ${w=2}$.
	One can note from the peripheries of exceptional rings in Fig.~\ref{fig4}(a) that the slopes at the rings are discontinuous. This brings dramatic consequences for the behavior of quasi-particles generated close to the ring.
	Figure~\ref{fig4}(b) shows the coalesced exceptional ring.
	
\begin{figure}[htbp]
		\centering
		\includegraphics[width=\columnwidth]{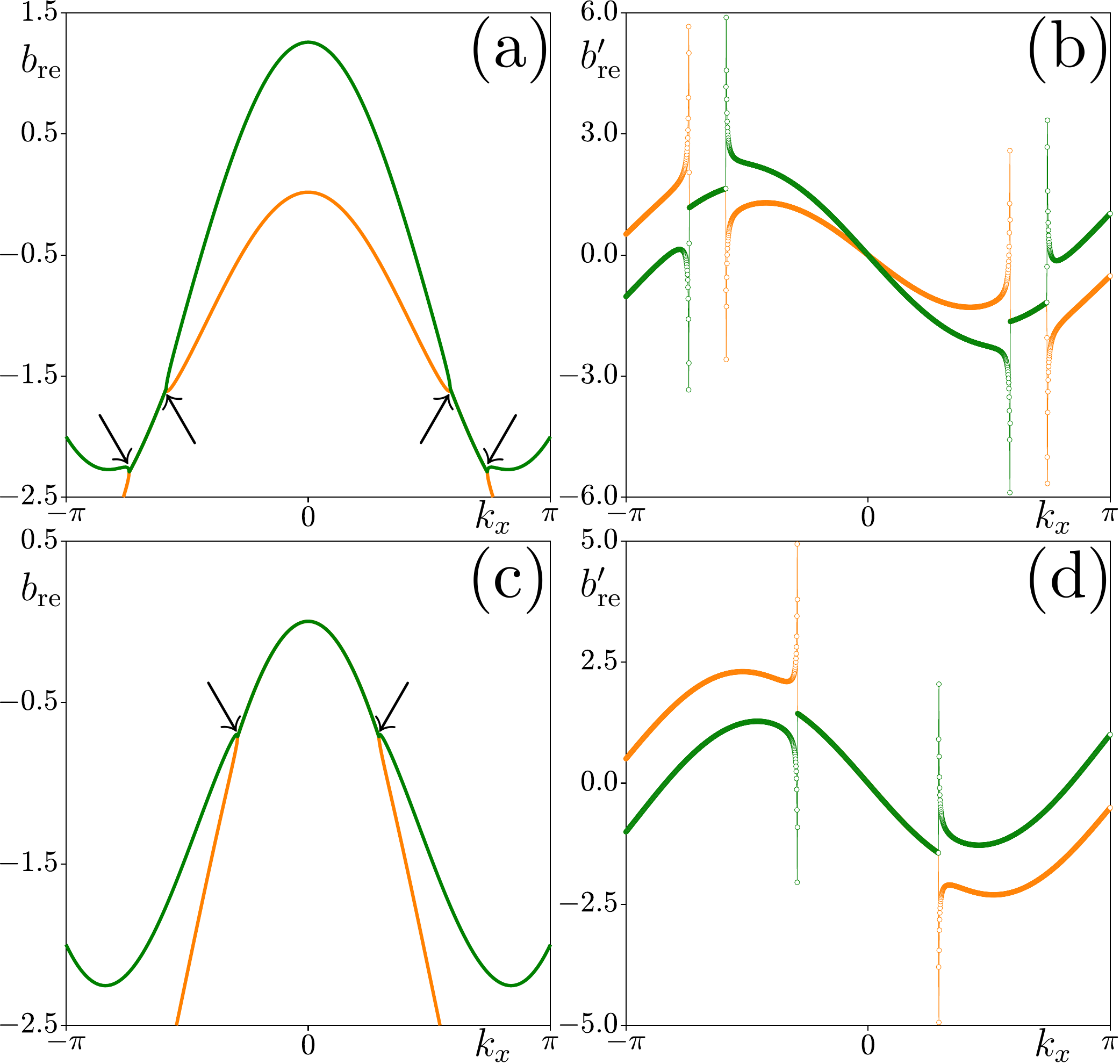}
		\caption{Exceptional rings in the cross-section ${k_y=0}$ and the corresponding first-order derivative ${b_{\rm re}'=-db_{\rm re}/dk_x}$. 
			(a,b) ${w=n=1}$. (c,d)  ${n=1}$ and ${w=2}$.
			Other parameters are same as those in Fig.~\ref{fig2}(b). 
			The color of the curve is in accordance with that of the corresponding band in Fig.~\ref{fig2}(b).
			Arrows in (a,c) point to the exceptional points.}
		\label{fig5}
	\end{figure}

	To better analyze the properties of type-II exceptional rings, we also display the band structure in the cross-section ${k_y=0}$ in Fig.~\ref{fig5}.
	There are four exceptional points in Fig.~\ref{fig5}(a), and at each exceptional point the dispersion exhibits a discontinuous slope of hyperbolic shape.
	Remembering that the gradient of the dispersion relation determines the group velocity via ${v=-b_{\rm re}'=db_{\rm re}/dk_x}$,
	we show the gradient of the bands in the cross-section ${k_y=0}$ in Fig.~\ref{fig5}(b).
	Clearly, the quasi-particle that is depicted by the wave packet will move with a rather large transverse group velocity when it is close to the exceptional point,
	since the velocity is approaching infinity at the point, and surely it will be far beyond the speed of the corresponding quasi-particles associated with type-II Dirac cones.
	These results are quite similar when the exceptional points coalesce, 
	as shown by the band structure in the cross-section ${k_y=0}$ of Fig.~\ref{fig5}(c), which exhibits two exceptional points
	and the corresponding gradient of the band structure in Fig.~\ref{fig5}(d).
	
	As discussed above, 
	the non-Hermitian lattice is ${\mathcal{PT}}$-symmetric if the intra-cell coupling $w$ is sufficiently stronger than the inter-cell coupling $n$.
	In the photonic lattice, one has to shift sites A and C closer to site B; 
	i.e., only the condition ${d_1=d_2<d}$ (rather than ${d_1=d_2>d}$) may restore the ${\mathcal{PT}}$ symmetry.
	Intensive numerical simulations demonstrate that the critical value of $p_{\rm im}$ is ${\sim0.571}$, at which the $\mathcal{PT}$ symmetry is broken. 
	In the \textbf{Supplemental Material}, we compare the light propagation dynamics in both conservative and ${\mathcal{PT}}$-symmetric lattices,
	with the same ${d_{1,2}=1.4}$, ${d=1.6}$, ${p_{\rm re}=10}$, and ${\sigma=0.5}$, according to Eq.~(\ref{eq1}).
	The loss in experiment can be controlled via the management of longitudinal landscapes or the quality of waveguide fabrication in photorefractive crystals or fused silica~\cite{weimann.nm.16.433.2016,xia.science.372.72.2021}.

	Summarizing, we have reported the non-Hermitian photonic lattice with natural type-II Dirac cones,
	by properly introducing the gain and loss to the sites in each unit cell.
	The exceptional rings associated with the type-II Dirac cones are no longer parallel with the Bloch wavevector plane.
	This kind of exceptional rings is different from the previously reported ones, and we call them the \textit{type-II exceptional rings}.
	We also find that the quasi-particles close to the type-II exceptional ring possess much higher velocities than those associated with the type-II Dirac cones.
	The results establish the link between the $\mathcal{PT}$ symmetry and the type-II Dirac cones,
	which may pave the way to develop active optical devices associated with the systems that violate the Lorentz invariance.
	
	\begin{backmatter}
		\bmsection{Funding}
		Natural Science Basic Research Program of Shaanxi Province, China (2024JC-JCQN-06);
		the National Natural Science Foundation of China (12304370, 12074308);
		Qatar National Research Fund (project NPRP13S-0121-200126);
		the Fundamental Research Funds for the Central Universities (sxzy012024146).
		
		\bmsection{Disclosures}
		The authors declare no conflicts of interest.

		\bmsection{Data Availability Statement}
		Data underlying the results presented in this paper are not publicly available at this time but may be obtained from the authors upon reasonable request.

		\bmsection{Supplemental document}
		See Supplement 1 for supporting content. 
		
	\end{backmatter}
	%
	%


\end{document}